\documentclass[12pt,psfig]{article}
\usepackage{amssymb}
\usepackage{amsmath,amsthm}
\usepackage{amsfonts}
\usepackage{graphicx}
\usepackage{graphics}
\numberwithin{equation}{section}

\usepackage{hyperref}

\begin{document}
\begin{center}\Large\textbf{Fractional D$p$-Branes
with Transverse and Tangential Dynamics in the Presence
of Background Fluxes}
\end{center}
\vspace{0.75cm}
\begin{center}{\large Shiva Heidarian and \large Davoud
Kamani}
\end{center}
\begin{center}
\textsl{\small{Physics Department, Amirkabir University of
Technology (Tehran Polytechnic)\\
P.O.Box: 15875-4413, Tehran, Iran\\
e-mails: kamani@aut.ac.ir ,
sh.Heidarian@aut.ac.ir\\}}
\end{center}
\vspace{0.5cm}

\begin{abstract}

We shall construct two boundary states which are corresponding to
a dynamical fractional
D$p$-brane in the presence of the fluxes of the Kalb-Ramond
field and a $U(1)$ gauge potential,
in the partially orbifold spacetime
$\mathbb{R}^{1, 5}\times\mathbb{C}^{2}/\mathbb{Z}_{2}$.
These states accurately describe the D$p$-brane in the
twisted and untwisted sectors under the orbifold projection.
We use them to compute the interaction of
two parallel fractional D$p$-branes with the
transverse velocities, tangential rotations and
tangential linear motions.
Various properties of the interaction, such as
its long-range force, will be discussed.

\end{abstract}

{\it PACS numbers}: 11.25.-w; 11.25.Uv

\textsl{\small{Keywords}}: Fractional D-branes;
Background fields; Dynamics; Boundary state formalism;
Twisted sector; Untwisted sector; Interaction.
\newpage
\section{Introduction}

D-branes represent important key roles in
understanding superstring theory and M-theory
\cite{1}-\cite{4}. They can be studied
through a powerful and adequate formalism, known
as the ``$boundary$ $state$'' formalism.
A boundary state is a physical state of closed string
that prominently encodes
all properties of a D-brane such as
tension, internal fields and dynamical
variables. This state elaborates that a
D-brane is a source for emitting all closed
string states. Therefore, overlap of two
boundary states via the closed string propagator
specifies the interaction amplitude of the
corresponding D-branes.
Hence, this strong procedure for
various configurations of the D-branes
has been vastly used \cite{5}-\cite{21}.

Among the various branes configurations
the fractional branes reveal profitable
behaviors \cite{22}-\cite{32}. For example, by a special
system of the fractional branes one can
demonstrate the gauge/gravity correspondence with
help of open/closed string duality \cite{27,32}.
In addition, the fractional branes drastically appear in the
various subjects of the string and M- theories.
For instance, the fractional branes
give some clues for defining
the Matrix theory \cite{33}-\cite{35}.

On one hand, we have the
dynamical branes which have a widespread
application in string theory. On the other hand,
the dressed branes, with background and internal fields,
exhibit various interesting properties.
For example, interactions of the branes
are accurately controlled by these fields.
Simultaneous consideration of the
dynamics, background fields, internal fields
and fractionality of the branes,
in the framework of superstring theory,
motivated and stimulated us to calculate the
boundary states and interaction of two
parallel fractional D$p$-branes
with the foregoing properties. Thus,
we shall consider the background field $B_{\mu\nu}$ and
two $U(1)$ internal potentials
$A^{(1,2)}_{\alpha}$ on the worldvolumes of the branes.
In this setup each brane has tangential
rotation, tangential and transverse linear motions.
The transverse dynamics of
the branes are along a non-orbifold perpendicular direction.
In the twisted superstring theory, via our orbifold,
the background spacetime partially
is non-compact orbifold with the
topological structure $\mathbb{R}^{1, 5}
\times\mathbb{C}^{2}/\mathbb{Z}_{2}$.
Finally, we shall separate the contribution of the
massless states of closed superstring
from the total interaction.
Our procedure will be the boundary state formalism.

This paper is organized as in the following.
In Sec. 2, we shall compute
the boundary states corresponding to a
dressed-fractional D$p$-brane
with tangential and transverse dynamics,
both for the untwisted and twisted sectors.
In Sec. 3, the interaction amplitude of two
parallel fractional D$p$-branes with the above
properties will be calculated.
In Sec. 4, we shall investigate the behavior of the
interaction for large distances of the branes
to obtain the long-range force of our setup.
Section 5 will be devoted to the conclusions.

\section{The boundary states associated with our D$p$-brane}
\hspace{0.5cm}
We consider a dressed-dynamical fractional
D$p$-brane which lives in the $d$-dimensional
spacetime and completely is transverse to the
non-compact orbifold $\mathbb{C}^{2}/\mathbb{Z}_{2}$.
The complex coordinates of $\mathbb{C}^{2}$
are constructed from $\{x^a|a= d-4, d-3, d-2, d-1\}$,
so that the $\mathbb{Z}_2$-group acts on them.
This group has the structure $\{e,h |h^2=e\}$,
in which under the action of the element
$h$ we have $x^a \rightarrow -x^a$.
The D$p$-brane has
stuck at the orbifold fixed-points,
which define a $(d-4)$-dimensional
hyperplane at the location $x^a=0$.
In the $d$-dimensional orbifoldized spacetime
the brane can possess the dimension $p\leq d-5$.

In order to construct the boundary state
of a dynamical fractional
D$p$-brane with background fields, we begin with the following
sigma-model action for closed string
\begin{eqnarray}
S=&-&\frac{1}{4\pi \alpha'}\int_{\Sigma}d^2\sigma
\left(\sqrt{-g}g^{rs}G_{\mu\nu}\partial_{r}X^{\mu}
\partial_s X^{\nu}+\epsilon^{rs}B_{\mu\nu}\partial_r X^{\mu}
\partial_s X^{\nu}\right)
\nonumber\\
&+&\frac{1}{2\pi \alpha'}\int_{\partial\Sigma}d\sigma
\left(A_{\alpha}\partial_{\sigma}X^{\alpha}+
\omega_{\alpha\beta}J^{\alpha\beta}_{\tau}\right),
\end{eqnarray}
where $B_{\mu\nu}$ is the Kalb-Ramond field,
and $A_\alpha(X)$ is a $U(1)$ gauge potential.
$g_{rs}$ and $G_{\mu\nu}$
are the metrics of the string worldsheet
and spacetime, respectively. The area
$\Sigma$ indicates the worldsheet of
the emitted closed string,
and $\partial\Sigma$ is its boundary.
The set $\{x^{\alpha}|\alpha=0,1,\ldots,p\}$
specifies the directions along the D$p$-brane worldvolume.

Here we assume the background fields
$G_{\mu\nu}$ and $B_{\mu\nu}$
to be constant, with
$G_{\mu\nu}=\eta_{\mu\nu}={\rm diag}(-1,1,\ldots,1 )$.
We use the conventional gauge
$A_{\alpha}=-\frac{1}{2}F_{\alpha \beta }X^{\beta}$
with a constant field strength $F_{\alpha \beta }$.
The brane's rotation-motion
term comprises a constant antisymmetric angular
velocity $\omega_{\alpha\beta}$.
The matrix elements $\omega_{0{\bar \alpha}}$ and
$\omega_{{\bar \alpha}{\bar \beta}}$, with
${\bar \alpha},{\bar \beta}\in \{1,2,\ldots,p\}$,
represent the linear
and angular velocities of the brane, respectively.
The variable $J^{\alpha \beta}_\tau=X^\alpha
\partial_\tau X^\beta-X^\beta \partial_\tau X^\alpha$
shows the angular momentum density.
Therefore, at present the dynamics of
the brane is inside its volume. We shall afterward
add a transverse motion too.
We should note that presence of the background field
$B_{\mu\nu}$ and the internal
field $A_\alpha$ gives rise to some preferred alignments inside
the brane worldvolume.
Thus, the Lorentz invariance in the worldvolume
has been explicitly broken. This clarifies that
the dynamics of the brane inside its volume is meaningful.

\subsection{The bosonic branch of the boundary state}

Vanishing the variation of the action
yields the following boundary state equations
for the twisted (T) and the untwisted (U) sectors,
via the orbifold projection,
\begin{eqnarray}
&~&\left[(\eta_{\alpha\beta}
+4\omega_{\alpha\beta})
\partial_{\tau}X^{\beta}
+\mathcal{F}_{\alpha\beta}
\partial_\sigma
X^{\beta}\right]_{\tau=0}
|B_x\rangle^{{\rm U}\backslash {\rm T}}=0~,
\nonumber\\
&~&\left(X^I-y^I\right)_{\tau=0}
|B_x\rangle^{{\rm U}\backslash {\rm T}}=0~,
\nonumber\\
&~&\left(X^a-y^a\right)_{\tau=0}
|B_x\rangle^{{\rm U}\backslash {\rm T}}=0~,
\end{eqnarray}
where $\mathcal{F}_{\alpha\beta}=
B_{\alpha\beta}-F_{\alpha\beta}$
is the total field strength.
The directions $\{x^I|I= p+1, \ldots, d-5\}$
will be used for both sectors.
In the twisted sector they refer
to the non-orbifold directions which are
perpendicular to the brane worldvolume.
For both sectors
the parameters $\{y^I|I= p+1, \ldots, d-5\}$
specify the position of the brane, and the
other position parameters,
due to the presence of the orbifold, for both
sectors are zero, i.e. $\{y^a =0 |a = d-4, \ldots,d-1\}$.
However, as we see there are mixed boundary conditions
along the brane worldvolume.

Now we introduce a transverse velocity to the brane.
Since the brane has stuck at the
orbifold fixed-points it cannot move along
the orbifold directions. Let the boost
direction be a member of the set
$\{x^{p+1}, \ldots, x^{d-5}\}$, which we call it
$x^{i_0}$. Hence,
Eqs. (2.2) under the boost find the features
\begin{eqnarray}
&~&[\partial_{\tau}(X^0-v^{i_0} X^{i_0})
+4\omega^{0}_{~~{\bar\beta}}
\partial_\tau X^{{\bar \beta}}
+\mathcal{F}^{0}_{~~{\bar\beta}}
\partial_\sigma X^{{\bar \beta}}]_{\tau=0}
|B_x\rangle^{{\rm U}\backslash {\rm T}}=0~,
\nonumber\\
&~&[\partial_{\tau}X^{\bar{\alpha}}
+4\gamma^2\omega^{\bar{\alpha}}_{~~0}
\partial_\tau (X^0-v^{i_0} X^{i_0})
+4\omega^{\bar{\alpha}}_{~~{\bar \beta}}
\partial_\tau X^{\bar \beta}
\nonumber\\
&~&+\gamma^2\mathcal{F}^{\bar{\alpha}}_{~~0}
\partial_\sigma (X^0-v^{i_0} X^{i_0})
+\mathcal{F}^{\bar{\alpha}}_{~~{\bar \beta}}
\partial_\sigma X^{\bar \beta}]_{\tau=0}
|B_x\rangle^{{\rm U}\backslash {\rm T}}=0~,
\nonumber\\
&~&(X^{i_0}-v^{i_0} X^{0}-y^{i_0})_{\tau=0}
|B_x\rangle^{{\rm U}\backslash {\rm T}}=0~,
\nonumber\\
&~&(X^i-y^i)_{\tau=0}
|B_x\rangle^{{\rm U}\backslash {\rm T}}=0~,
\nonumber\\
&~&(X^a)_{\tau=0}|B_x\rangle^{{\rm U}\backslash {\rm T}}=0~,
\end{eqnarray}
where $\gamma =1/\sqrt{1-(v^{i_0})^2}$,
and $i\in \{p+1,\ldots,{\hat {i_0}}, \ldots,d-5\}$, i.e.
$i \neq {i_0}$. In order to have a transverse
motion the dimension of the brane is restricted by
$p \leq d-6$.

For each sector, by applying the mode
expansion of the closed string coordinates into
Eqs. (2.3), the boundary state equations will be written in
terms of the string oscillators.
For the twisted sector the mode expansions of the
closed string coordinates
$X^\alpha$, $X^{i_0}$ and $X^i$ possess the common form
\begin{equation}
X^\rho(\sigma,\tau)=x^\rho+2\alpha'p^\rho
\tau+\frac{i}{2}\sqrt{2\alpha'}
\sum_{m\neq0}\frac{1}{m}
\left(\alpha_m^{\rho}
e^{-2im(\tau-\sigma)}+\tilde{\alpha}_m^\rho
e^{-2im(\tau+\sigma)}\right)~,\;\rho \in \{\alpha, {i_0}, i\},
\end{equation}
and the string coordinates along the orbifold have the feature
\begin{eqnarray}
X^a(\sigma,\tau)=\frac{i}{2}\sqrt{2\alpha'}
\sum_{r\in\mathbb{Z}+1/2}
\frac{1}{r}\left(\alpha_r^{a}
e^{-2ir(\tau-\sigma)}+\tilde{\alpha}_r^a
e^{-2ir(\tau+\sigma)}\right).
\end{eqnarray}
For the untwisted sector the mode expansion
of the string coordinates are as the conventional cases.

On the basis of the mode expansions we obtain
\begin{eqnarray}
&~&\bigg{[}\bigg{(}\gamma(\delta^{0}_{~~\lambda}
-v^{i_0}\delta^{{i_0}}_{~~\lambda})
+\gamma(4\omega^{0}_{~~{\bar\alpha}}
-\mathcal{F}^{0}_{~~\bar{\alpha}})
\delta^{\bar{\alpha}}_{~~\lambda}\bigg{)}
\alpha_{m}^{\lambda}+
\nonumber\\
&~&+\bigg{(}\gamma(\delta^{0}_{~~\lambda}
-v^{i_0}\delta^{{i_0}}_{~~\lambda})
+\gamma(4\omega^{0}_{~~{\bar\alpha}}
+\mathcal{F}^{0}_{~~\bar{\alpha}})
\delta^{\bar{\alpha}}_{~~\lambda}\bigg{)}
\tilde{\alpha}_{-m}^{\lambda}
\bigg{]}|B^{(\rm osc)}_x\rangle^{{\rm U}\backslash {\rm T}}=0~,
\nonumber\\
\nonumber\\
&~&\bigg{[}\left(\delta^{\bar{\alpha}}_{~~\lambda}+
\gamma^{2}(4\omega^{{\bar\alpha}}_{~~0}
-\mathcal{F}^{\bar{\alpha}}_{~~0})
(\delta^{0}_{~~\lambda}
-v^{i_0}\delta^{{i_0}}_{~~\lambda})+
(4\omega^{{\bar\alpha}}_{~~\bar{\beta}}
-\mathcal{F}^{\bar{\alpha}}_{~~\bar{\beta}})
\delta^{\bar{\beta}}_{~~\lambda}
\right)\alpha_{m}^{\lambda}+
\nonumber\\
&~&+\left( \delta^{\bar{\alpha}}_{~~\lambda}+
\gamma^{2}(4\omega^{{\bar\alpha}}_{~~0}
+\mathcal{F}^{\bar{\alpha}}_{~~0})
(\delta^{0}_{~~\lambda}
-v^{i_0}\delta^{{i_0}}_{~~\lambda})+
(4\omega^{{\bar\alpha}}_{~~\bar{\beta}}
+\mathcal{F}^{\bar{\alpha}}_{~~\bar{\beta}})
\delta^{\bar{\beta}}_{~~\lambda}\right)
\tilde{\alpha}_{-m}^{\lambda}
\bigg{]}|B^{(\rm osc)}_x\rangle^{{\rm U}\backslash {\rm T}}=0~,
\nonumber\\
\nonumber\\
&~&\left[(\delta^{{i_0}}_{~~\lambda}
-v^{i_0}\delta^{0}_{~~\lambda})\alpha_{m}^{\lambda}
-(\delta^{{i_0}}_{~~\lambda}
-v^{i_0}\delta^{0}_{~~\lambda})
\tilde{\alpha}_{-m}^{\lambda}\right]
|B^{(\rm osc)}_x\rangle^{{\rm U}\backslash {\rm T}}=0,~
\nonumber\\
&~&(\alpha_{m}^{i}-\tilde{\alpha}_{-m}^{i})
|B^{(\rm osc)}_x\rangle^{{\rm U}\backslash {\rm T}}=0,~
\nonumber\\
&~&(\alpha_{r}^{a}-\tilde{\alpha}_{-r}^{a})
|B^{(\rm osc)}_x\rangle^{{\rm U}\backslash {\rm T}}=0,~
\end{eqnarray}
where $\lambda \in \{\alpha,{i_0}\}$.
For the zero-mode parts of both sectors Eqs. (2.3) yield
\begin{eqnarray}
&~&\left[p^{0}-v^{{i_0}}p^{{i_0}}+4\omega_{~~\bar{\beta}}^{0}
~p^{\bar{\beta}}
\right] |B^{(0)}_x\rangle^{{\rm U}\backslash {\rm T}}=0~,
\nonumber\\
&~&\left[ p^{\bar{\alpha}}
+4\gamma^{2}\omega_{~~0}^{\bar{\alpha}}
(p^{0}-v^{{i_0}}p^{{i_0}})
+4\omega_{~~\bar{\beta}}^{\bar{\alpha}}~
p^{\bar{\beta}}\right]
|B^{(0)}_x\rangle^{{\rm U}\backslash {\rm T}}=0~,
\nonumber\\
&~&(x^{i_0}-v^{i_0} x^0 -y^{i_0})|B^{(0)}_x
\rangle^{{\rm U}\backslash {\rm T}}=0~,
\nonumber\\
&~&(x^{i}-y^{i})|B^{(0)}_x\rangle^{{\rm U}\backslash {\rm T}}=0~,
\nonumber\\
&~&(x^{a})|B^{(0)}_x\rangle^{\rm U}=0~,
\end{eqnarray}
where we exerted the decomposition
$|B_x\rangle^{{\rm U}\backslash {\rm T}}=
|B^{(0)}_x\rangle^{{\rm U}\backslash {\rm T}}
\otimes|B^{(\rm osc)}_x\rangle^{{\rm U}\backslash {\rm T}}$.

According to the first two equations of Eqs. (2.7) we receive
the following relations among the eigenvalues of the
momentum components
\begin{eqnarray}
&~&p^{0}-v^{{i_0}}p^{{i_0}}=-4
\omega^{0}_{~\bar{\beta}}~p^{\bar{\beta}},
\nonumber\\
&~&\Omega^{\bar{\alpha}}_{~\bar{\beta}}
~p^{\bar{\beta}}=0,
\nonumber\\
&~&\Omega^{\bar{\alpha}}_{~\bar{\beta}} \equiv
\delta^{\bar{\alpha}}_{~\bar{\beta}}
-16\gamma^{2}\omega^{\bar{\alpha}}_{~0}
~\omega^{0}_{~\bar{\beta}}
+4\omega^{\bar{\alpha}}_{~\bar{\beta}}~.
\end{eqnarray}
We observe that the tangential dynamics of the brane
connects the momentum components of the emitted closed string.
For the third relation we have two choices: if the
$p\times p$ matrix
$\Omega^{\bar{\alpha}}_{~\bar{\beta}}$
is invertible, i.e.
$\det\Omega^{\bar{\alpha}}_{~\bar{\beta}}\neq0 $,
then all $p^{\bar{\alpha}}$s
must vanish, and hence $p^0$ and $p^{i_0}$
identically become zero. The second choice is
$\det\Omega^{\bar{\alpha}}_{~\bar{\beta}}=0 $,
which is a constraint between the
$1+p(p+1)/2$ parameters $\{\omega_{\alpha\beta} , v^{i_0}\}$.
In this case, $p^{\bar{\alpha}}$s can be nonzero, so
$p^0$ and $p^{i_0}$ can also be nonzero.
This non-vanishing momentum extremely is different from the
usual case in which
the closed strings are emitted perpendicular to
the brane. The nonzero momentum implies that the brane dynamics
effectively induces a peculiar potential
on the emitted closed strings.
For example, the second choice for
the D2-brane eventuates to the following equation
between the tangential, normal and angular velocities
of the brane
\begin{eqnarray}
(1+16\omega^2_{12})(1-v^2_\perp)-16v^2_{\parallel}=0,
\end{eqnarray}
where $v_\perp = v^{i_0}$ and
$v^2_{\parallel} = \omega^2_{01}+\omega^2_{02}$.
For simplification of the
calculations we shall apply the first choice.

The coherent state method
elaborates the following solutions for the oscillating parts
of the boundary state
\begin{eqnarray}
|B^{(\rm osc)}_x\rangle^{\rm T}&=&
\sqrt{\det Q}\;
\exp\left[{-\sum_{m=1}^{\infty}
\left(\frac{1}{m}
\alpha_{-m}^{\rho}S_{\rho\rho'}
\tilde{\alpha}_{-m}^{\rho'}
\right)}\right]
\nonumber\\
&\times&\exp\left[\sum_{r=1/2}^{\infty}
\left(\frac{1}{r}
\alpha_{-r}^{a}\tilde{\alpha}_{-r}^{a}\right)\right]
|0\rangle_\alpha
\otimes|0\rangle_{\tilde{\alpha}}~,
\nonumber\\
\nonumber\\
|B^{(\rm osc)}_x\rangle^{\rm U}&=&
\sqrt{\det Q}\;
\exp\left[{-\sum_{m=1}^{\infty}
\left(\frac{1}{m}
\alpha_{-m}^{\mu}\tilde{S}_{\mu\nu}
\tilde{\alpha}_{-m}^{\nu}\right)}\right]
|0\rangle_\alpha
\otimes|0\rangle_{\tilde{\alpha}}~,\label{aos}
\end{eqnarray}
where $\rho,\rho'\in\{\alpha,i,{i_0}\}$
and $\mu,\nu\in\{\alpha, a, i, {i_0}\}$.
The matrices $S$ and $\tilde{S}$ are defined by
\begin{eqnarray}
S_{\rho\rho'}&=&\left((Q^{-1}N)_{\lambda\lambda'},
-\delta_{ij}\right)~,
\nonumber\\
\tilde{S}_{\mu\nu}&=&\left((Q^{-1}N)_{\lambda\lambda'},
-\delta_{ij}, -\delta_{ab}\right)~,
\nonumber\\
\nonumber\\
Q_{~~\lambda}^{0}&=&
\gamma(\delta^{0}_{~~\lambda}
-v^{i_0}\delta^{{i_0}}_{~~\lambda})
+\gamma(4\omega^{0}_{~~{\bar\alpha}}
-\mathcal{F}^{0}_{~~\bar{\alpha}})
\delta^{\bar{\alpha}}_{~~\lambda}~,
\nonumber\\
Q_{~~\lambda}^{\bar{\alpha}}&=&
\delta^{\bar{\alpha}}_{~~\lambda}+
\gamma^{2}(4\omega^{{\bar\alpha}}_{~~0}
-\mathcal{F}^{\bar{\alpha}}_{~~0})
(\delta^{0}_{~~\lambda}
-v^{i_0}\delta^{{i_0}}_{~~\lambda})+
(4\omega^{{\bar\alpha}}_{~~\bar{\beta}}
-\mathcal{F}^{\bar{\alpha}}_{~~\bar{\beta}})
\delta^{\bar{\beta}}_{~~\lambda}~,
\nonumber\\
Q_{~~\lambda}^{{i_0}}&=&\delta^{{i_0}}_{~~\lambda}
-v^{i_0}\delta^{0}_{~~\lambda}~,
\nonumber\\
\nonumber\\
N_{~~\lambda}^{0}&=&
\gamma(\delta^{0}_{~~\lambda}
-v^{i_0}\delta^{{i_0}}_{~~\lambda})
+\gamma(4\omega^{0}_{~~{\bar\alpha}}
+\mathcal{F}^{0}_{~~\bar{\alpha}})
\delta^{\bar{\alpha}}_{~~\lambda}~,
\nonumber\\
N_{~~\lambda}^{\bar{\alpha}}&=&
\delta^{\bar{\alpha}}_{~~\lambda}+
\gamma^{2}(4\omega^{{\bar\alpha}}_{~~0}
+\mathcal{F}^{\bar{\alpha}}_{~~0})
(\delta^{0}_{~~\lambda}-
v^{i_0}\delta^{{i_0}}_{~~\lambda})+
(4\omega^{{\bar\alpha}}_{~~\bar{\beta}}
+\mathcal{F}^{\bar{\alpha}}_{~~\bar{\beta}})
\delta^{\bar{\beta}}_{~~\lambda}~,
\nonumber\\
N_{~~\lambda}^{{i_0}}&=&-\delta^{{i_0}}_{~~\lambda}
+v^{i_0}\delta^{0}_{~~\lambda}~.
\end{eqnarray}
The normalization factors in Eqs. (2.10)
can be deduced from the disk partition function
\cite{36}-\cite{38}.
Precisely, the quadratic form of the
tangential dynamics term, accompanied by the
gauge $A_\alpha = -\frac{1}{2}F_{\alpha\beta}X^\beta$,
induces a quadratic form for the boundary portion of the action
(2.1). Therefore, path integration on this Gaussian action
manifestly introduces the prefactor
$\prod_{n=1}^\infty \left( \det Q\right)^{-1}$
to Eqs. (2.10). Using the regularization
$\prod_{n=1}^\infty a \longrightarrow 1/\sqrt{a}$
the prefactors find the above square root feature.

Note that the coherent state method
gives the boundary states (2.10)
under the conditions $S S^{T}=\mathbf{1}$ and
${\tilde S} {\tilde S}^{T}=\mathbf{1}$.
These equations eventuate to the following
relations among the variables
$\{\omega_{\alpha\beta},F_{\alpha\beta},B_{\alpha\beta} ,v^{i_0}\}$,
\begin{eqnarray}
&~& \omega^{0}_{~~\bar{\alpha}}
~\mathcal{F}^{0}_{~~\bar{\alpha}}=0,
\nonumber\\
&~&~\mathcal{F}^{\bar{\alpha}}_{~~\bar{\beta}}
\;\omega^{0 \bar{\beta}}
+~\omega^{\bar{\alpha}}_{~~\bar{\beta}}\;
\mathcal{F}^{0 \bar{\beta}}=0,
\nonumber\\
&~&\mathcal{F}^{\bar{\alpha}}_{~~\bar{\kappa}}
~\omega^{\bar{\beta} \bar{\kappa}}+
\mathcal{F}^{\bar{\beta}}_{~~\bar{\kappa}}
~\omega^{\bar{\alpha} \bar{\kappa}}-
\gamma^{2}\left( \omega^{\bar{\alpha}}_{~~0}
~\mathcal{F}^{\bar{\beta}}_{~~0}+
\omega^{\bar{\beta}}_{~~0}
~\mathcal{F}^{\bar{\alpha}}_{~~0}\right) =0~.
\end{eqnarray}
Thus, from the total $d-p-5+3p(p+1)/2$ parameters
of each brane, i.e.
$\{\omega_{\alpha\beta},F_{\alpha\beta},B_{\alpha\beta} ,
v^{i_0}, y^i\}$, only $p^2+d-p-6$ of them remain independent.

Making use of the commutation relation
$[x^\mu , p^\nu ]=i\eta^{\mu \nu}$
the zero-mode parts of the boundary states find the solutions
\begin{eqnarray}
|B^{(0)}_x\rangle^{\rm T} &=& \frac{T_p}{2}
~\delta \left({x}^{{i_0}}-v^{{i_0}}x^{0}
-y^{{i_0}}\right)|p^{i_0}=0\rangle
 \prod_{i} \left[ \delta
\left({x}^{i}-y^{i}\right)
|p^{i}=0\rangle \right]
\prod_{\alpha}|p^{\alpha}=0\rangle,
\nonumber\\
|B^{(0)}_x\rangle^{\rm U} &=& \frac{T_p}{2}
~\delta\left({x}^{{i_0}}
-v^{{i_0}}x^{0}-y^{{i_0}}\right)|p^{i_0}=0\rangle
 \prod_{i}\left[ \delta \left({x}^{i}
-y^{i}\right)|p^{i}=0\rangle
\right]
\nonumber\\
&\times& \prod_{a}\left[ \delta \left(x^a
\right)|p^a=0\rangle \right]
\prod_{\alpha}|p^{\alpha}=0\rangle.
\end{eqnarray}
The constant factor $T_p$ is the tension of the D$p$-brane.

In the bosonic string theory the total boundary states,
corresponding to the two sectors, are given by
$$|B_x\rangle^{{\rm U}\backslash {\rm T}}=
|B_x^{\rm (osc)}\rangle^{{\rm U}\backslash {\rm T}}
\otimes|B^{(0)}_x\rangle^{{\rm U}\backslash {\rm T}}
\otimes|B_{\rm gh}\rangle~,$$
where $|B_{\rm gh}\rangle$ is the known boundary
state, associated with the conformal ghosts,
and for both sectors obviously is the same.

\subsection{The fermionic branch of the boundary state}

The worldsheet supersymmetry implies that we can exert the
following replacements on the bosonic boundary state
equations (2.3) to extract their fermionic counterparts
\begin{eqnarray}
\partial_{+}X^\mu(\sigma ,\tau)&\rightarrow&
-i\eta\psi_{+}^\mu(\tau +\sigma)~,
\nonumber\\
\partial_{-}X^\mu(\sigma,\tau)&\rightarrow&
-\psi_{-}^\mu(\tau - \sigma)~,
\label{jk}
\end{eqnarray}
where $\partial_\pm = (\partial_\tau \pm \partial_\sigma)/2$,
and $\eta=\pm1$ is saved for the GSO projection on the boundary
states.

Similar to the bosonic part of the boundary state
the fermionic part also includes
the twisted and untwisted sectors.
For constructing the boundary state equations in terms
of the fermionic oscillators
we use the fermionic mode expansions
of each sector. For the untwisted sector
the worldsheet fields $\psi^\mu_\pm$
have the well-known mode expansions, and for the twisted
sector they have the following expansions
\begin{eqnarray}
&~& \psi_{+}^\mu=\sum_{t}\tilde{\psi}_{t}^\mu
e^{-2it(\tau+\sigma)}~,
\nonumber\\
&~& \psi_{-}^\mu=\sum_{t}\psi_{t}^\mu e^{-2it(\tau-\sigma)}~,
\end{eqnarray}
where in the twisted NS-NS sector there are
\begin{eqnarray}
&~& \psi_{t}^\rho~\text{and}~\tilde{\psi}_{t}^\rho,
~~t\in \mathbb{Z}+1/2,
\nonumber\\
&~& \psi_{r}^a~\text{and}~\tilde{\psi}_{r}^a,~~r\in \mathbb{Z},
\nonumber
\end{eqnarray}
and in the twisted R-R sector we have
\begin{eqnarray}
&~& \psi_{t}^\rho~\text{and}~\tilde{\psi}_{t}^\rho,
~~t\in \mathbb{Z},\nonumber\\
&~& \psi_{r}^a~\text{and}~\tilde{\psi}_{r}^a,~~r\in
\mathbb{Z}+1/2.
\nonumber
\end{eqnarray}
The indices ``$\rho$'' and ``$a$'' indicate
the non-orbifold and orbifold directions, respectively.
Since for the superstring theory the critical dimension is $d=10$,
we select the sets $\{a|a=6,7,8,9\}$
and $\{\rho |\rho =0,1,2,3,4,5\}$
for the orbifold and non-orbifold directions, respectively.

Introducing the replacements (2.14) into
Eqs. (2.3) and using the above mode expansions, we obtain
\begin{eqnarray}
&~&\bigg{[}\left(\gamma(\delta^{0}_{~~\lambda}
-v^{i_0}\delta^{{i_0}}_{~~\lambda})
+\gamma(4\omega^{0}_{~~{\bar\alpha}}
-\mathcal{F}^{0}_{~~\bar{\alpha}})
\delta^{\bar{\alpha}}_{~~\lambda}\right)
\psi_{t}^{\lambda}
\nonumber\\
&~&-i\eta\left(\gamma(\delta^{0}_{~~\lambda}
-v^{i_0}\delta^{{i_0}}_{~~\lambda})
+\gamma(4\omega^{0}_{~~{\bar\alpha}}
+\mathcal{F}^{0}_{~~\bar{\alpha}})
\delta^{\bar{\alpha}}_{~~\lambda}\right)
\tilde{\psi}_{-t}^{\lambda}
\bigg{]}|B_{\psi},
\eta\rangle^{{\rm U}\backslash {\rm T}}=0~,
\nonumber\\
\nonumber\\
&~&\bigg{[}\left(\delta^{\bar{\alpha}}_{~~\lambda}+
\gamma^{2}(4\omega^{{\bar\alpha}}_{~~0}
-\mathcal{F}^{\bar{\alpha}}_{~~0})
(\delta^{0}_{~~\lambda}
-v^{i_0}\delta^{{i_0}}_{~~\lambda})+
(4\omega^{{\bar\alpha}}_{~~\bar{\beta}}
-\mathcal{F}^{\bar{\alpha}}_{~~\bar{\beta}})
\delta^{\bar{\beta}}_{~~\lambda}
\right)\psi_{t}^{\lambda}
\nonumber\\
&~&-i\eta\left(\delta^{\bar{\alpha}}_{~~\lambda}+
\gamma^{2}(4\omega^{{\bar\alpha}}_{~~0}
+\mathcal{F}^{\bar{\alpha}}_{~~0})
(\delta^{0}_{~~\lambda}-v^{i_0}\delta^{{i_0}}_{~~\lambda})+
(4\omega^{{\bar\alpha}}_{~~\bar{\beta}}
+\mathcal{F}^{\bar{\alpha}}_{~~\bar{\beta}})
\delta^{\bar{\beta}}_{~~\lambda}\right)
\tilde{\psi}_{-t}^{\lambda}
\bigg{]}|B_{\psi};\eta\rangle^{{\rm U}\backslash {\rm T}}=0~,
\nonumber\\
\nonumber\\
&~&\left[(\delta^{{i_0}}_{~~\lambda}
-v^{i_0}\delta^{0}_{~~\lambda})\psi_{t}^{\lambda}
+i\eta(\delta^{{i_0}}_{~~\lambda}
-v^{i_0}\delta^{0}_{~~\lambda})
\tilde{\psi}_{-t}^{\lambda}\right]
|B_{\psi};\eta\rangle^{{\rm U}\backslash {\rm T}}=0,~
\nonumber\\
&~&(\psi_{t}^{i}+i\eta \tilde{\psi}_{-t}^{i})
|B_{\psi};\eta\rangle^{{\rm U}\backslash {\rm T}}=0,~
\nonumber\\
&~&(\psi_{r}^{a}+ i\eta \tilde{\psi}_{-r}^{a})
|B_{\psi};\eta\rangle^{{\rm U}\backslash {\rm T}}=0.
\end{eqnarray}
Now let decompose
$|B_{\psi};\eta\rangle^{{\rm U}\backslash {\rm T}}=
|B^{\rm (osc)}_{\psi};\eta\rangle^{{\rm U}\backslash {\rm T}}
\otimes|B^{(0)}_{\psi};\eta\rangle^{{\rm U}\backslash {\rm T}}$.
Thus. the oscillating part of these equations can be rewritten
in the compact forms
\begin{eqnarray}
&~&\left(\psi_t^\rho-i\eta
S^\rho_{~~\rho'}\tilde{\psi}^{\rho'}_{-t}\right)
|B_\psi^{\rm (osc)};\eta\rangle^{\rm T}=0~,
\nonumber\\
&~&(\psi_{r}^{a}+ i\eta \tilde{\psi}_{-r}^{a})
|B_{\psi}^{\rm (osc)};\eta\rangle^{\rm T}=0 ~,
\nonumber\\
&~&\left(\psi_t^\mu-i\eta\tilde{S} ^\mu_{~~\nu}
\tilde{\psi}^{\nu}_{-t}\right)
|B_\psi^{\rm (osc)};\eta\rangle^{\rm U}=0~,
\end{eqnarray}
where for the first and third equations
$t\in \mathbb{Z}-\{0\}\;(t\in \mathbb{Z}+1/2)$ is related
to the R-R (NS-NS) sector, and in the
second equation there is $r\in \mathbb{Z}+1/2 \;
(r\in \mathbb{Z}-\{0\})$ for the R-R (NS-NS) sector.

The zero-mode parts of Eqs. (2.16), for the
NS-NS and R-R sectors, take the features
\begin{eqnarray}
&~& \left(\psi_0^\rho-i\eta S^\rho_{\;\;\;\rho'}
\tilde{\psi}^{\rho'}_{0}\right){|B_\psi^{(0)},
\eta\rangle^{\rm T}_{\rm R}}=0~,\label{qw}
\nonumber\\
&~& \left(\psi_0^a+i\eta~
\tilde{\psi}^a_{0}\right){
|B_\psi^{(0)};\eta\rangle^{\rm T}_{\rm NS}}=0~,
\nonumber\\
&~& \left(\psi_0^\mu-i\eta \tilde{S}^\mu_{\;\;\;\nu}
\tilde{\psi}^{\nu}_{0}\right){|B_\psi^{(0)},
\eta\rangle^{\rm U}_{\rm R}}=0~.
\end{eqnarray}
Note that, unlike the untwisted sector, the NS-NS
portion of the twisted sector also possesses a
zero-mode part which originates from the orbifold directions.

\subsubsection{The boundary states of the NS-NS sectors}

On the basis of the coherent state
method the oscillating parts of
the NS-NS boundary states for both sectors are given by
\begin{eqnarray}
&~&|B_\psi;\eta\rangle^{\rm T}_{\rm NS}
=\exp\left[i\eta\sum_{t=1/2}^{\infty}\psi_{-t}^\rho~
S_{\rho\rho'}~\tilde{\psi}_{-t}^{\rho'}\right]
\exp\left[i\eta\sum_{r=1}^{\infty}\psi_{-r}^a
~\tilde{\psi}_{-r}^a\right]{|B_\psi^{(0)};\eta
\rangle^{\rm T}_{\rm NS}}~,
\nonumber\\
&~& |B_\psi;\eta\rangle^{\rm U}_{\rm NS}
=\exp\left[i\eta\sum_{t=1/2}^{\infty}\psi_{-t}^\mu~
\tilde{S}_{\mu\nu}~\tilde{\psi}_{-t}^{\nu}\right]
{|0\rangle^{{\rm U}}_{\rm NS}}~.
\end{eqnarray}
The second equation of Eqs. (2.18) elucidates that
the state $|B_\psi^{(0)};\eta \rangle^{\rm T}_{\rm NS}$
is independent of the background fields and
dynamics of the brane. It has the solution
\cite{26}, \cite{29},
\begin{eqnarray}
{|B_\psi^{(0)};\eta\rangle^{\rm T}_{\rm NS}}=\left(\bar{C}\;
\frac{1+i\eta\bar{\Gamma}}{1+i\eta}\right)_{LM}~
|L\rangle\otimes|\tilde{M}\rangle~,
\end{eqnarray}
where $\bar{C}$ is the charge conjugation matrix of the
group $SO(4)$,
$\bar{\Gamma}={\hat{\Gamma}}^6 {\hat{\Gamma}}^7
{\hat{\Gamma}}^8{\hat{\Gamma}}^9$,
and
$|L\rangle$ and $|\tilde{M}\rangle$ are spinor
states of $SO(4)$.
\subsubsection{The boundary states of the R-R sectors}

By solving Eqs. (2.17) we acquire the following solutions
for the oscillating parts of the R-R boundary states
\begin{eqnarray}
|B_\psi;\eta\rangle^{\rm T}_{\rm R}
&=&\frac{1}{\sqrt{\det Q}}
\exp\left[i\eta\sum_{t=1}^{\infty}\psi_{-t}^\rho~
S_{\rho\rho'}~\tilde{\psi}_{-t}^{\rho'}\right]
\nonumber\\
&\times& \exp\left[i\eta\sum_{r=1/2}^{\infty}
\psi_{-r}^a~\tilde{\psi}_{-r}^a\right]
{|B_\psi^{(0)};\eta\rangle^{\rm T}_{\rm R}}~,
\nonumber\\
\nonumber\\
|B_\psi;\eta\rangle^{\rm U}_{\rm R}
&=&\frac{1}{\sqrt{\det Q}}
~\exp\left[i\eta\sum_{t=1}^{\infty}\psi_{-t}^\mu~
\tilde S_{\mu\nu}~\tilde{\psi}_{-t}^{\nu}\right]
{|B_\psi^{(0)};\eta\rangle^{\rm U}_{\rm R}}~.
\end{eqnarray}
The reversed determinants, in
contrast with the bosonic part, i.e. Eqs. (2.10), is due to the
Grassmannian nature of the fermionic variables.

The zero-mode boundary states
$|B_\psi^{(0)};\eta\rangle^{\rm T}_{\rm R}$
and $|B_\psi^{(0)};\eta\rangle^{\rm U}_{\rm R}$
are the solutions of the first and third
equations of (2.18). The explicit forms of them are given by
\begin{eqnarray}
&~& |B_\psi^{(0)};\eta\rangle^{\rm T}_{\rm R}
=\gamma\left(\tilde{C}(\Gamma'^{0}
+v^{{i_0}}\Gamma'^{{i_0}})\;\Gamma'^1\ldots\Gamma'^p~
\frac{1+i\eta\tilde{\Gamma}}
{1+i\eta}~G'\right)_{A'B'}~
|A'\rangle \otimes|\tilde{B'}\rangle~,
\nonumber\\
&~& |B_\psi^{(0)};\eta\rangle^{\rm U}_{\rm R}
=\gamma\left({C}(\Gamma^{0}
+v^{{i_0}}\Gamma^{{i_0}})\;\Gamma^1\ldots\Gamma^p~
\frac{1+i\eta{\Gamma^{11}}}{1+i\eta}~G\right)_{AB}~
|A\rangle \otimes|\tilde{B}\rangle~.
\end{eqnarray}
In the twisted sector ${\tilde C}$ is the charge conjugate
matrix of the group $SO(1, 5)$,
the $\Gamma'$-matrices
satisfy the Clifford algebra in the
six dimensions and
$\tilde{\Gamma}={\Gamma'}^0 {\Gamma'}^1\ldots
{\Gamma'}^5$, then $|A'\rangle$ and
$|\tilde{B'}\rangle$ are spinors of
$SO(1,5)$. In the untwisted sector
$C$ is the charge conjugate matrix of $SO(1, 9)$, and
$|A\rangle$ and $|\tilde{B}\rangle$ are spinors of this group.
The $32 \times 32$ and $8 \times 8$ matrices
$G$ and $G'$ satisfy the following equations
\begin{eqnarray}
&~&\Gamma'^\lambda G' -M^\lambda _{~~\lambda'}
G'\Gamma'^{\lambda'}
-v^{{i_0}}\Gamma'^{{i_0}}\Gamma'^\lambda \Gamma'^{0}G'
-v^{{i_0}}\Gamma'^{{i_0}}\Gamma'^{0}M^\lambda _{~~\lambda'}
G'\Gamma'^{\lambda'}=0,
\nonumber\\
&~&\Gamma^\lambda G-M^\lambda _{~~\lambda'}
G\Gamma^{\lambda'}
-v^{{i_0}}\Gamma^{{i_0}}\Gamma^\lambda \Gamma^{0}G
-v^{{i_0}}\Gamma^{{i_0}}\Gamma^{0}M^\lambda _{~~\lambda'}
G \Gamma^{\lambda '}=0,
\end{eqnarray}
where $M^\lambda _{~~\lambda'}=(Q^{-1}N)^{\lambda}_{~~\lambda'}$.
Using the algebra of the Dirac matrices
these equations can be rewritten in the suitable forms
\begin{eqnarray}
&~&\Gamma'^\lambda(G'+v^{{i_0}}\Gamma'^{{i_0}}\Gamma'^{0}G')
-M^\lambda _{\;\;\;\lambda '}(G'
+v^{{i_0}}\Gamma'^{{i_0}}\Gamma'^{0}G')\Gamma'^{\lambda'}
=2v^{{i_0}}\eta^{{i_0}\lambda }\Gamma'^{0}G'~,
\nonumber\\
&~&\Gamma^\lambda(G
+v^{{i_0}}\Gamma^{{i_0}}\Gamma^{0}G)
-M^\lambda_{\;\;\;\lambda '}
(G+v^{{i_0}}\Gamma^{{i_0}}\Gamma^{0}G)\Gamma^{\lambda'}
=2v^{{i_0}}\eta^{{i_0}\lambda}\Gamma^{0}G~.
\end{eqnarray}
Therefore, $G'$ and $G$ explicitly find the solutions
\begin{eqnarray}
G'&=&\left[(1+v^{{i_0}}\Gamma'^{{i_0}}
\Gamma'^{0})-2
v^{i_0}\Gamma'^{{i_0}}\Gamma'^{0}(1
+R^{i_0}_{~~\lambda'}\Gamma'^{i_0}
\Gamma'^{\lambda'})^{-1}\right]^{-1}
\nonumber\\
&\times& :\exp\left(\frac{1}{2}{\bar{\Phi}}_{\lambda\lambda'}
\Gamma'^\lambda\Gamma'^{\lambda'}\right):~,
\nonumber\\
G &=& \left[ (1+v^{{i_0}}\Gamma^{i_0}
\Gamma^{0})-2
v^{i_0}\Gamma^{{i_0}}\Gamma^{0}(1
+R^{i_0}_{~~\lambda'}\Gamma^{i_0}
\Gamma^{\lambda'})^{-1}\right]^{-1}
\nonumber\\
&\times&
:\exp\left(\frac{1}{2}{\bar{\Phi}}_{\lambda\lambda'}
\Gamma^\lambda\Gamma^{\lambda'}\right):~,
\nonumber\\
\bar{\Phi}&=&(\Phi-\Phi^T)/2\;,
\nonumber\\
\Phi_{\lambda\lambda'} & \equiv &
\left((PM+1)^{-1}(PM-1)\right) _{\lambda\lambda'}~,~~~~~~~
\;\label{rt}
\end{eqnarray}
where $R=PM$, and the matrix $P$ is defined by
$P^{\lambda}_{~~\lambda'}=(\delta^{\alpha}
_{~~\beta}\;,-\delta^{i_0}_{~~{i_0}})$ with
$P^{\alpha}_{~~i_0}=0$.
The conventional notation $:~:$ implies that
we must expand the exponentials
with the convention that all Dirac matrices anticommute,
hence only a finite number of terms remain.
In the absence of the dynamical variables we have
$\bar\Phi_{\alpha\beta}=\mathcal{F}_{\alpha\beta}$
which is in accordance with the
conventional results of the literature.

For instance, the antisymmetric matrix
$\bar\Phi_{\lambda \lambda'}$ corresponding to
a dressed fractional D2-brane, which is parallel to the
$x^1x^2$-plane and its tangential velocity is
along the $x^3$-direction, has the following structure
\begin{eqnarray}
\bar{\Phi}_{\lambda \lambda}&=&~0 ,
\nonumber\\
\bar{\Phi}_{01}&=&\bigg{[}
2(1-v^2)\left(\mathcal{F}_{02}\omega_{12}
+\mathcal{F}_{12}(4\omega_{01}\omega_{12}-\omega_{02})
\right)-8\mathcal{F}_{02}\omega_{01}\omega_{02}
\nonumber\\
&~&-\mathcal{F}_{01} \left((1-v^2)(1+8\omega_{12}^{2})
-8\omega_{02}^{2}\right)\bigg{]}/W ,
\nonumber\\
\bar{\Phi}_{02}&=&\bigg{[}
-2(1-v^2)\left(\mathcal{F}_{01}\omega_{12}
-\mathcal{F}_{12}(4\omega_{02}\omega_{12}+\omega_{01})
\right)-8\mathcal{F}_{01}\omega_{01}\omega_{02}
\nonumber\\
&~&-\mathcal{F}_{02} \left((1-v^2)(1+8\omega_{12}^{2})
-8\omega_{01}^{2}\right)\bigg{]}/W ,
\nonumber\\
\bar{\Phi}_{03}&=&~v,
\nonumber\\
\bar{\Phi}_{12}&=&\bigg{[}-2\left(
\mathcal{F}_{01}\omega_{02}-\mathcal{F}_{02}\omega_{01}
\right)
-8\left(\mathcal{F}_{01}\omega_{01}\omega_{12}
+\mathcal{F}_{02}\omega_{02}\omega_{12}\right)
\nonumber\\
&~&-\mathcal{F}_{12}\left(1-v^{2}
-8(\omega_{01}^{2}+\omega_{02}^{2})\right)
\bigg{]}/W ,
\nonumber\\
\bar{\Phi}_{13}&=&~0 ,
\nonumber\\
\bar{\Phi}_{23}&=&~0 ,
\nonumber\\
W &\equiv & (1-v^2)\left(16\omega^2_{12}-1
\right)+16\left(\omega_{01}^{2}+\omega_{02}^{2}
\right) .
\end{eqnarray}

\subsection{The total boundary states}

For eliminating the closed string tachyon and
preserving the supersymmetry
we should apply the GSO projection.
The total GSO-projected boundary state
is a linear combination of two states with
$\eta =\pm 1$. Thus, the total physical boundary states in
the twisted and untwisted sectors are given by
\begin{eqnarray}
&~&|B\rangle^{\rm T}_{\rm NS,R}=\frac{1}{2}
\left(|B,+\rangle^{\rm T}_{\rm NS,R}
+|B,-\rangle^{\rm T}_{\rm NS,R}\right)~,
\nonumber\\
&~&|B\rangle^{\rm U}_{\rm NS}=\frac{1}{2}
\left(|B,+\rangle^{\rm U}_{\rm NS}
-|B,-\rangle^{\rm U}_{\rm NS}\right)~,
\nonumber\\
&~&|B\rangle^{\rm U}_{\rm R}=\frac{1}{2}
\left(|B,+\rangle^{\rm U}_{\rm R}
+|B,-\rangle^{\rm U}_{\rm R}\right)~,\label{dft}
\end{eqnarray}
where the states
$|B;\eta\rangle^{\rm {\rm U}\backslash {\rm T}}_{\rm NS,R}$
are defined by the partial states
$$|B;\eta\rangle^{\rm  {\rm U}\backslash {\rm T}}_{\rm NS,R}=
|B_x\rangle^{\rm {\rm U}\backslash {\rm T}}\otimes
|B_\psi;\eta\rangle^{\rm {\rm U}\backslash {\rm T}}_{\rm NS,R}
\otimes|B_{\rm gh}\rangle
\otimes|B_{\rm sgh};\eta\rangle_{\rm NS,R}~.$$
As we see the GSO-projection selects
different combinations in the
NS-NS and R-R portions of the untwisted sector
while for the twisted sector,
because of the orbifold projection,
it chooses similar structure
for both NS-NS and R-R parts.

Note that the ghost and superghost boundary states
$|B_{\rm gh}\rangle$ and $|B_{\rm sgh};\eta\rangle_{\rm NS,R}$
are not influenced by the dynamics of the brane,
the orbifold projection and the background fields.
For the next purposes we bring in the boundary state
of the R-R super-conformal ghosts
\begin{eqnarray}
|B_{\rm sgh} ; \eta\rangle_{\rm R}
&=& \exp \bigg{[} i\eta \sum_{n=1}^\infty \left(
\gamma_{-n}{\tilde \beta}_{-n}-
\beta_{-n}{\tilde \gamma}_{-n} \right)
\nonumber\\
&+& i\eta \gamma_0 {\tilde \beta}_0 \bigg{]}
|P=-1/2\;,\;{\tilde P}=-3/2 \rangle ,
\end{eqnarray}
where the vacuum of the superghosts is in the picture
$(-1/2 , -3/2)$, and it is annihilated by
${\tilde \gamma}_0$ and $\beta_0$.

\section{The D-branes interaction}

The D-branes interactions
have extensively appeared in the main subjects of physics.
For example, the following phenomena
have satisfactory descriptions via the D-branes interactions:
the gauge/gravity correspondence \cite{27,32},
presence of the dark matter \cite{39},
extra gravity inside our universe
\cite{40,41}, origin of the inflation \cite{42,43},
creation of the Big-Bang by the D-branes collision \cite{44},
and may other physical phenomena, e.g. see \cite{45, 46, 47}.

The boundary state formalism allows us to directly compute
the cylinder amplitude in the closed
string channel. The ends of the cylinder lie on
the D-branes and represent the boundaries
of the closed superstring worldsheet.
This superstring is emitted by one of the branes and
then is absorbed by the other one. Since
each D-brane couples to the all closed superstring states via its
corresponding boundary state,
it obviously is a source for procreating any closed superstring states.
The interaction amplitude is calculated as the
tree-level diagram between
the two boundary states, associated with
the D-branes. Therefore, two
D-branes prominently interact through the exchange
of closed superstrings.

The orbifold projection imposes two parts
for the interaction: one
part due to the untwisted sector and another
portion from the twisted sector
\begin{eqnarray}
&~&\mathcal{A}^{\rm Total}=\mathcal{A}^{\rm U}+\mathcal{A}^{\rm T},
\nonumber\\
&~& \mathcal{A}^{{\rm U}\backslash {\rm T}}=
{}_{\rm NS}^{{\rm U}\backslash {\rm T}}\langle  B_1
|D^{{\rm U}\backslash {\rm T}}_{\rm NS}
|B_2 \rangle^{{\rm U}\backslash {\rm T}}_{\rm NS}
+{}_{\rm R}^{{\rm U}\backslash {\rm T}}\langle  B_1
|D^{{\rm U}\backslash {\rm T}}_{\rm R}
|B_2 \rangle^{{\rm U}\backslash {\rm T}}_{\rm R},
\nonumber\\
&~& D^{{\rm U}\backslash {\rm T}}_{\rm NS,R}
=2\alpha'\int_{0}^{\infty}
dt~e^{-tH_{\rm NS,R}^{{\rm U}\backslash {\rm T}}}~,
\end{eqnarray}
where $D^{{\rm U}\backslash {\rm T}}$
and $H^{{\rm U}\backslash {\rm T}}$ are the closed
superstring propagators and Hamiltonians, respectively.
One should apply the GSO-projected boundary
states from Eqs. (2.27). The
Eqs. (3.1) clarify that in the interaction all possible
forces between two D-branes,
which are exchanged by the R-R and NS-NS states of closed
superstring, have been taken into account.
However, the total Hamiltonians are given by
\begin{eqnarray}
&~& H^{\rm T\backslash U}_{\rm NS}=
H_x^{\rm T\backslash U}+H_{\psi,{\rm NS}}^{\rm T\backslash U}
+H_{\rm gh}+H_{\rm sgh, NS},
\nonumber\\
&~& H^{\rm T\backslash U}_{\rm R}=
H_x^{\rm T\backslash U}+H_{\psi,{\rm R}}^{\rm T\backslash U}
+H_{\rm gh}+H_{\rm sgh, R}.
\nonumber
\end{eqnarray}
For the twisted sector we have
\begin{eqnarray}
&~& H_x^{\rm T}=\alpha'p^{\rho}p_{\rho}
+2\sum_{n=1}^{\infty}(\alpha_{-n}^{\rho}
\alpha_{n\rho}
+\tilde{\alpha}_{-n}^{\rho}
\tilde{\alpha}_{n\rho})
+2\sum_{r=1/2}^{\infty}
(\alpha_{-r}^{a}\alpha_{ra}
+\tilde{\alpha}_{-r}^{a}
\tilde{\alpha}_{ra})-\frac{2}{3}~,
\nonumber\\
&~& H_{\psi,{\rm NS}}^{\rm T}=
2\sum_{t=1/2}^{\infty}(t\psi_{-t}^{\rho}
\psi_{t\rho}
+t\tilde{\psi}_{-t}^{\rho}
\tilde{\psi}_{t\rho})
+2\sum_{r=1}^{\infty}
(r\psi_{-r}^{a}\psi_{ra}
+r\tilde{\psi}_{-r}^{a}
\tilde{\psi}_{ra})+\frac{1}{6}~,
\nonumber\\
&~& H_{\psi,{\rm R}}^{\rm T}=
2\sum_{t=1}^{\infty}(t\psi_{-t}^{\rho}
\psi_{t\rho}
+t\tilde{\psi}_{-t}^{\rho}\tilde{\psi}_{t\rho})
+2\sum_{r=1/2}^{\infty}(r\psi_{-r}^{a}\psi_{ra}
+r\tilde{\psi}_{-r}^{a}
\tilde{\psi}_{ra})+\frac{2}{3}~.
\end{eqnarray}
Note that, by enumerating the
ghosts and superghosts contributions,
the zero-point energies of the total Hamiltonians
$H_{\rm NS}^{\rm T}$ and $H_{\rm R}^{\rm T}$ vanish.
The untwisted sector comprises the following total Hamiltonians
\begin{eqnarray}
H_{{\rm NS}}^{\rm U}&=& H_{\rm gh}+H_{\rm sgh, NS}
+\alpha'p^{\mu}p_{\mu}
+2\sum_{n=1}^{\infty}(\alpha_{-n}^{\mu}
\alpha_{n\mu}+\tilde{\alpha}_{-n}^{\mu}
\tilde{\alpha}_{n\mu})
\nonumber\\
&+&2\sum_{t=1/2}^{\infty}(t\psi_{-t}^{\mu}
\psi_{t\mu}+t\tilde{\psi}_{-t}^{\mu}
\tilde{\psi}_{t\mu})-\frac{5}{2},
\nonumber\\
H_{{\rm R}}^{\rm U}&=&H_{\rm gh}+H_{\rm sgh, R}
+\alpha'p^{\mu}p_{\mu}
+2\sum_{n=1}^{\infty}(\alpha_{-n}^{\mu}
\alpha_{n\mu}+\tilde{\alpha}_{-n}^{\mu}
\tilde{\alpha}_{n\mu})
\nonumber\\
&+& 2\sum_{t=1}^{\infty}(t\psi_{-t}^{\mu}
\psi_{t\mu}+t\tilde{\psi}_{-t}^{\mu}
\tilde{\psi}_{t\mu}).
\end{eqnarray}
The total Hamiltonian $H_{{\rm NS}}^{\rm U}$
possesses nonzero vacuum energy, while the zero-point
energy of the total Hamiltonian $H_{{\rm R}}^{\rm U}$
vanishes.

\subsection{Partial amplitudes}

Now we study the various parts of the total interaction amplitude
for two parallel dynamical fractional
D$p$-brane with background fields. The branes completely sit
at the orbifold fixed-points with the
transverse velocities $v_{1}^{{i_0}}$ and $v_{2}^{{i_0}}$
along the non-orbifold perpendicular direction $x^{i_0}$,
the tangential dynamics $\omega_{(1)\alpha\beta}$ and
$\omega_{(2)\alpha\beta}$, the internal field strength
$F_{(1)\alpha\beta}$ and $F_{(2)\alpha\beta}$,
and the background field $B_{\mu\nu}$.

\subsubsection{The amplitude in the untwisted sector}

Since the untwisted sector is not affected by
the orbifold projection here we merely give the final result
for its amplitude
\begin{eqnarray}
\mathcal{A}^{\rm U}&=&
\frac{T_{p}^{2}\alpha'V_p}
{8(2\pi)^{8-p}}~\frac{1}{\mathcal{V}}
~\int_{0}^{\infty}dt
~\Big{(}\sqrt{\frac{\pi}{\alpha' t}}\Big{)}^{8-p}
~\exp\left(-\frac{1}{4\alpha't}\sum_{i}
{\left(y_{1}^{i}-y_{2}^{i}\right)^2}\right)
\nonumber\\
&\times&\bigg\{~\frac{1}{q}\sqrt{\det (Q^{\dag}_{1}Q_{2})}
~\bigg{[}
~\prod_{n=1}^{\infty}\frac{\det\left(1+M^T_{1}
M_{2}~q^{2n-1}\right)}{\det\left(1-M^T_{1}M_{2}~q^{2n}
\right)}\left(\frac{1-q^{2n}}
{1+q^{2n-1}}\right)^{p-6}
\nonumber\\
&-&\prod_{n=1}^{\infty}\frac{\det\left(1-M^T_{1}
M_{2}~q^{2n-1}\right)}{\det\left(1-M^T_{1}M_{2}~q^{2n}
\right)}\left(\dfrac{1-q^{2n}}
{1-q^{2n-1}}\right)^{p-6}\bigg{]}
\nonumber\\
&+& \xi~\prod_{n=1}^{\infty}\frac{\det\left(1+M^T_{1}
M_{2}~q^{2n}\right)}{\det\left(1-M^T_{1}M_{2}~q^{2n}
\right)}\left(\dfrac{1-q^{2n}}
{1+q^{2n}}\right)^{p-6}+\xi'~ \bigg\}~,
\label{po}
\end{eqnarray}
where $q=e^{-2t}$, $\mathcal{V}=|v_{1}^{{i_0}}-v_{2}^{{i_0}}|$
is the relative velocity of the branes,
$V_p$ is the volume
of each brane, and $\xi$ and $\xi'$ are defined by
\begin{eqnarray}
&~&\xi\equiv \frac{1}{2}
{\rm Tr}\left(C ~G_{2}~C^{-1}G_{1}^{T}
\left[-1+v_{1}^{{i_0}}v_{2}^{{i_0}}
+(v_{1}^{{i_0}}-v_{2}^{{i_0}})(\Gamma^{0})^{T}
(\Gamma^{{i_0}})^{T}\right]
\right) ~,
\nonumber\\
&~&\xi'\equiv {\rm Tr}\left(
C ~G_{2}~C^{-1}G_{1}^{T}\left[
-1+v_{1}^{{i_0}}v_{2}^{{i_0}}
+(v_{1}^{{i_0}}-v_{2}^{{i_0}})(\Gamma^{0})^{T}
(\Gamma^{{i_0}})^{T}\right]\Gamma^{11}
\right) ~.
\end{eqnarray}
The four terms in Eq. (3.4) originate from the NS-NS,
NS-NS$(-1)^{F}$, R-R and R-R$(-1)^{F}$.
In fact, the factor $\xi'$ for the usual configurations
of the branes, e.g. stationary and bare of the internal
and background fields, vanishes. In other words,
some special setups, such as our setup,
receive nonzero values for $\xi'$. In Eq. (3.4)
the three possible signs of $(\xi, \xi')$
reveal the interactions of the brane-brane,
antibrane-antibrane and brane-antibrane systems.
The exponential factor shows
the damping nature of the interaction
with respect to the square distance
of the branes. This damping factor will also appear
in the twisted sector.

\subsubsection{The amplitude in the twisted NS-NS sector}

Applying the total GSO-projected boundary state
of the twisted NS-NS sector we acquire the following
partial amplitude
\begin{eqnarray}
\mathcal{A}^{\rm T}_{\rm NS-NS}(\eta_1 , \eta_2)
&=&\frac{T_{p}^{2}\alpha'V_p}
{4(2\pi)^{4-p}}~\frac{1}{\mathcal{V}}
\sqrt{\det (Q^{\dag}_{1}Q_{2})}
\int_{0}^{\infty}dt \bigg\{
~\Big{(}\sqrt{\frac{\pi}{\alpha't}}\Big{)}^{4-p}
\nonumber\\
&\times& \delta_{\eta_1 \eta_2 ,1}~
\exp\left(-\frac{1}{4\alpha't}\sum_{i}
{\left(y_{1}^{i}-y_{2}^{i}\right)^2}\right)
\nonumber\\
&\times& \prod_{n=1}^{\infty}\left[
\frac{\det\left(1+M^T_{1}
M_{2}~q^{2n-1}\right)}{\det\left(1-M^T_{1}M_{2}~q^{2n}
\right)}\left(
\frac{1-q^{2n}}{1+q^{2n-1}}\right)^{p-2}
\left(\frac{1+q^{2n}}{1-q^{2n-1}}\right)^{4}\right]
\bigg\}.
\nonumber\\
\end{eqnarray}
The six factors in the infinite product come from
the oscillators, and have the following origins.
The determinants of the numerator and denominator
respectively are the effects of
the fermions and bosons, along the directions of
the worldvolumes and transverse velocities.
The exponent of the factor
$\prod{}_{n=1}^{\infty}(1-q^{2n})^{p-2}$
is $2 + (p-4)$, where +2 is the ghosts
contribution and $p-4$ is for the bosons along the
non-orbifold perpendicular directions except
the transverse velocity direction. The exponent of the product
$\prod{}_{n=1}^{\infty}(1+q^{2n-1})^{2-p}$
is $-2 + (4-p)$, where -2 is the superghosts
contribution and $4-p$ turns up from the
fermions along the
non-orbifold perpendicular directions except
the transverse velocity direction.
The factors $\prod{}_{n=1}^{\infty}(1+q^{2n})^{4}$
and $\prod{}_{n=1}^{\infty}(1-q^{2n-1})^{-4}$
arise from the fermions and bosons in the
orbifold directions, respectively.

We observed that the zero-mode part of the boundary state
in the twisted NS-NS sector, i.e. Eq. (2.20), has
a non-trivial structure.
Hence, the spin structure NS-NS$(-1)^{F}$ does
not contribute to the interaction.
\subsubsection{The amplitude in the twisted R-R sector}

In the calculation of the interaction
amplitude of the twisted R-R sector we receive a
contribution from the zero-modes of
the super-conformal ghosts which is divergent, i.e.,
according to the Eq. (2.28) we obtain
\begin{eqnarray}
&~&{}_{\rm R}\langle B^{(0)}_{\rm sgh};\eta_1
|B^{(0)}_{\rm sgh};\eta_2\rangle_{\rm R}
=\sum_{n=0}^\infty (-\eta_1\eta_2)^n ,
\nonumber
\end{eqnarray}
where for $\eta_1\eta_2 =+1$ is an alternating series,
and for $\eta_1\eta_2 =-1$ becomes divergent.
Hence, it requires a suitable regularization.
Similar to the Refs.\cite{7,48} we insert the regulator
$\mathcal R(x) = x^{2G_0}$ as follows
\begin{eqnarray}
&~&{}^{\rm T}_{\rm R}\langle B^{(0)}_1;\eta_1
|B^{(0)}_2;\eta_2\rangle_{\rm R}^{\rm T}
\equiv \lim_{x\rightarrow1}
{}^{\rm T}_{\rm R}\langle B^{(0)}_{1};
\eta_1|\mathcal R(x)|B^{(0)}_{2};
\eta_2\rangle_{\rm R}^{\rm T}
\nonumber\\
&~&=\lim_{x\rightarrow1}
\bigg{[}{}_{\rm R}
\langle B^{(0)}_{\rm sgh};\eta_1|
x^{2G_0}|B^{(0)}_{\rm sgh};
\eta_2\rangle_{\rm R}\times
{}^{\rm T}_{\rm R}\langle B^{(0)}_{1\psi};
\eta_1|B^{(0)}_{2\psi};
\eta_2\rangle_{\rm R}^{\rm T}\bigg{]},
\end{eqnarray}
where $G_0=-\gamma_0\beta_0$.
The superghost factor eventuates to the result
$1/(1+\eta_1\eta_2 x^2)$. Thus, by the following
insertion of $\beta_0$, $\gamma_0$, ${\tilde \beta}_0$
and ${\tilde \gamma}_0$ in the superghost part of Eq. (3.7)
for $\eta_1=-\eta_2 \equiv \eta$ we acquire
\begin{eqnarray}
\lim_{x\rightarrow1}
{}_{\rm R}\langle B^{(0)}_{\rm sgh};
\eta|x^{2G_0}\delta \left(\beta_{0}
-\frac{1}{4\pi}\gamma_0 \right)
\delta \left({\tilde \beta_{0}} +\frac{1}{4\pi}
{\tilde \gamma_0} \right)|B^{(0)}_{\rm sgh};
-\eta\rangle_{\rm R}=1~.
\end{eqnarray}
For $\eta_1=-\eta_2$
this defines a regular amplitude in the twisted R-R sector.
Therefore, we receive
\begin{eqnarray}
\lim_{x\rightarrow1}
{}^{\rm T}_{\rm R}\langle
B^{(0)}_{1};\eta_1|\mathcal R(x)
|B^{(0)}_{2};\eta_2\rangle_{\rm R}^{\rm T} &=&{\rm Tr}
\bigg{\{}\bigg[\tilde{C}\;(\Gamma'^0
+v^{{i_0}}_{2}\Gamma'^{{i_0}})\Gamma'^{1}\ldots \Gamma'^p
\frac{1+i\eta_2\tilde{\Gamma}}{1+i\eta_2}
G'_2\bigg]\tilde{C}^{-1}
\nonumber\\
&\times& \bigg{[}\tilde{C}\;(\Gamma'^0
+v^{{i_0}}_{1}\Gamma'^{{i_0}})\Gamma'^{1}\ldots
\Gamma'^p G'_1
\frac{1+i\eta_1\tilde{\Gamma}}{1-i\eta_1}
\bigg{]}^{\rm T}\tilde{C}^{-1}\bigg{\}}
\nonumber\\
&\equiv &\tilde{\xi}\;\delta_{\eta_1\eta_2,1}
+\tilde{\xi'}\;\delta_{\eta_1\eta_2,-1}
\end{eqnarray}
where $\tilde{\xi}$ and $\tilde{\xi'}$ have
the definitions
\begin{eqnarray}
\tilde{\xi}&\equiv &
\frac{1}{2}{\rm Tr}\left(
\tilde{C}~G'_{2}~\tilde{C}^{-1}
(G'_{1})^{T}\left[
-1+v_{1}^{{i_0}}v_{2}^{{i_0}}+(v_{1}^{{i_0}}
-v_{2}^{{i_0}})(\Gamma'^{0})^{T}
(\Gamma'^{{i_0}})^{T}\right]\right),
\nonumber\\
\tilde{\xi}'&\equiv & {\rm Tr}\left(
\tilde{C}~G'_{2}~\tilde{C}^{-1}
(G'_{1})^{T}\left[
-1+v_{1}^{{i_0}}v_{2}^{{i_0}}+(v_{1}^{{i_0}}
-v_{2}^{{i_0}})(\Gamma'^{0})^{T}
(\Gamma'^{{i_0}})^{T}\right]\tilde{\Gamma}\right).
\end{eqnarray}

In fact, this regularization has been also used in
the untwisted R-R sector with
the $\Gamma$-matrices of the group $SO(1, 9)$. Since
the superghost boundary state is independent of
the orbifold projection we applied
the result (3.8) to the untwisted R-R sector.

Adding all these together the partial amplitude
of the twisted R-R sector is accomplished as
\begin{eqnarray}
\mathcal{A}^{\rm T}_{\rm R-R}(\eta_1 , \eta_2)&=&
\frac{T_{p}^{2}\alpha'V_p}
{16(2\pi)^{4-p}}~\dfrac{1}{\mathcal{V}}
~\int_{0}^{\infty}dt
~\Big{(}\sqrt{\frac{\pi}{\alpha' t}}\Big{)}^{4-p}
\nonumber\\
&\times& \exp\left(-\frac{1}{4\alpha't}\sum_{i}
{\left(y_{1}^{i}-y_{2}^{i}\right)^2}\right)
\nonumber\\
&\times& \bigg{[}~\delta_{\eta_1 \eta_2 , 1}\;\tilde{\xi}
\prod_{n=1}^{\infty}\frac{\det\left(1+M^T_{1}
M_{2}~q^{2n}\right)}{\det\left(1-M^T_{1}M_{2}~q^{2n}\right)}
\left(\dfrac{1-q^{2n}}{1+q^{2n}}\right)^{p-2}
\left(\dfrac{1+q^{2n-1}}{1-q^{2n-1}}\right)^{4}
\nonumber\\
&+& \delta_{\eta_1 \eta_2 , -1}\;\tilde{\xi}'~\bigg{]}.
\label{po}
\end{eqnarray}
These two terms are
corresponding to the R-R and R-R$(-1)^{F}$ spin structures,
respectively. The
R-R$(-1)^{F}$ portion merely receives a non-vanishing
contribution from the fermionic zero-modes.
The origins of the six factors in the infinite
product in Eq. (3.11) are analogous
to the description after Eq. (3.6),
in which here the fermions and super-conformal
ghosts live in the R-R sector.
Similar to the untwisted R-R sector,
in Eq. (3.11) the normalizing factors of
the fermions and bosons exactly cancel each other.

\subsection{The total interaction amplitude}

As we said the total interaction amplitude
possesses two main parts
$\mathcal{A}^{\rm Total}=\mathcal{A}^{\rm U}+\mathcal{A}^{\rm T}$.
The untwisted part was exhibited by Eq. (3.4).
The twisted part is specified by the following summation
\begin{eqnarray}
&~& \mathcal{A}^{\rm T}
=\mathcal{A}^{\rm T}(+ , +)
+\mathcal{A}^{\rm T}(+ , -)
+\mathcal{A}^{\rm T}(- , +)
+\mathcal{A}^{\rm T}(- , -)~,
\nonumber\\
&~& \mathcal{A}^{\rm T}(\eta_1 , \eta_2)
=\mathcal{A}_{\rm NS-NS}^{\rm T}(\eta_1 , \eta_2)
+\mathcal{A}_{\rm R-R}^{\rm T}(\eta_1 , \eta_2),
\end{eqnarray}
where $\eta_1 , \eta_2 \in \{+1,-1\}$.
The signs of the quantities ${\tilde \xi}$ and
${\tilde \xi'}$ in the amplitude (3.12)
indicate the interactions of the brane-brane,
antibrane-antibrane and brane-antibrane systems.

By comparing the amplitudes of the untwisted
sector $\mathcal{A}^{\rm U}$
and twisted sector $\mathcal{A}^{\rm T}$ we note that
presence of the orbifold directions drastically
induces significant effects on the interaction.
However, presence of the various parameters in the setup,
i.e., the matrix elements of the $U(1)$ field strengths
and the Kalb-Ramond tensor, the tangential and transverse
and angular velocities of the branes,
the dimension of the branes, the coordinates
of the branes locations, and the orbifold
effects inspires a general feature
to the total interaction amplitude $\mathcal{A}^{\rm Total}$.
The strength of the interaction is accurately
controlled by these parameters.

In fact, the relative transverse velocity of the branes
generally breaks the supersymmetry. Therefore,
our setup does not preserve
enough value of the supersymmetry, and hence it does not satisfy
the BPS no-force condition. This can be manifestly seen
by the fact that, for the D$p$-branes with the same angular
velocity $\omega_{(1)\alpha\beta}=\omega_{(2)\alpha\beta}$
and identical internal fields,
the attraction force of the NS-NS states
is not compensated by the repulsive force of the R-R states.

Note that since the ghost and superghost
parts of the boundary states completely are independent of the
background fields, the branes dynamics and
orbifold projection, their contributions have
been introduced by manipulation into the
partial amplitudes (3.4), (3.6) and (3.11).
\section{Interaction of the branes with large distance}

In each interaction theory behavior of the
associated amplitude
after an enough long time represents a reliable long-range force
of the theory. Thus, for the interacting distant branes the
massless closed superstring states
extremely possess a dominant contribution
to the interaction, while the contribution of all massive states,
except the tachyon state, vanish. The long-range amplitude
$\mathcal{A}^{(0)}$ is obtained by taking the limit
$t\rightarrow\infty$ of the oscillating portions of the
total amplitude $\mathcal{A}^{\rm Total}$.
Note that the superstring states are merely
defined by the oscillators. Hence,
since the other time dependent factors come
from the bosonic zero-modes we shall not take the limit of them.

The total interaction amplitude of the distant branes is
\begin{eqnarray}
\mathcal{A}^{\rm (0)Total}=\mathcal{A}^{\rm (0)U}+
\mathcal{A}^{\rm (0)T}_{\rm NS-NS}+
\mathcal{A}^{\rm (0)T}_{\rm R-R}~,
\end{eqnarray}
where the untwisted part of this amplitude is given by
\begin{eqnarray}
\mathcal{A}^{\rm (0)U}&=&
\frac{T_{p}^{2}\alpha'V_p}
{8(2\pi)^{8-p}}~\frac{1}{\mathcal{V}}\;
\int_{0}^{\infty}dt~\bigg\{
\Big{(}\sqrt{\frac{\pi}{\alpha't}}\Big{)}^{8-p}
\exp\left(-\frac{1}{4\alpha't}\sum_{i}
{\left(y_{1}^{i}-y_{2}^{i}\right)^2}\right)\bigg\}
\nonumber\\
&\times& \bigg{(}\sqrt{\det (Q^\dagger_1 Q_2)}
\bigg{[}12-2p+2{\rm Tr}(M^{T}_{1}M_{2})\bigg{]}
+\xi +\xi' \bigg{)}~.
\end{eqnarray}

According to Eq. (3.6),
for computing the contribution of the twisted NS-NS sector,
we must apply the limit
\begin{eqnarray}
&~&\lim_{t \to \infty}
\prod_{n=1}^{\infty}\left[
\frac{\det\left(1+M^T_{1}
M_{2}~q^{2n-1}\right)}{\det\left(1-M^T_{1}M_{2}~q^{2n}
\right)}\left(
\frac{1-q^{2n}}{1+q^{2n-1}}\right)^{p-2}
\left(\frac{1+q^{2n}}{1-q^{2n-1}}\right)^{4}\right]
\nonumber\\
&~&\longrightarrow \left(1+
\left[6-p+{\rm Tr}(M^{T}_{1}M_{2})\right]
e^{-2t} \right)~.
\label{po}
\end{eqnarray}
Therefore, we acquire the partial amplitude
\begin{eqnarray}
\mathcal{A}^{\rm (0)T}_{\rm NS-NS}&=&
\frac{T_{p}^{2}\alpha'V_p}
{2(2\pi)^{4-p}}~\frac{1}{\mathcal{V}}\;
\sqrt{\det(Q^\dagger_1 Q_2)}
\int_{0}^{\infty}dt~\bigg\{
\Big{(}\sqrt{\frac{\pi}{\alpha' t}}\Big{)}^{4-p}
\nonumber\\
&\times& \exp\left(-\frac{1}{4\alpha't}\sum_{i}
{\left(y_{1}^{i}-y_{2}^{i}\right)^2}\right)
\nonumber\\
&\times& \lim_{t \to \infty}\left(1+
\left[6-p+{\rm Tr}(M^{T}_{1}M_{2})\right]
e^{-2t}\right)\bigg\}~.
\label{po}
\end{eqnarray}
The brackets in the last lines of Eqs. (4.2) and (4.4)
reveal the valuable contribution of the graviton, dilaton
and Kalb-Ramond states to the long-range force.
As we see the contribution of these states
in the twisted sector is exponentially damped.
This is because of the fact that
under the orbifold projection these states become massive.
This effect was done by deforming the zero-point energy of
the corresponding Hamiltonian.
In fact, due to this modified zero-point energy,
the ground state of closed superstring is changed to a
massless state.
Hence, the long-range force (4.4) completely
originates from the exchange of this massless state.

Making use of Eq. (3.11),
for the twisted R-R sector we should exert the limit
\begin{eqnarray}
&~&\lim_{t\rightarrow \infty}
\prod_{n=1}^{\infty}\left[
\frac{\det\left(1+M^T_{1}
M_{2}~q^{2n}\right)}{\det\left(1-M^T_{1}M_{2}~q^{2n}\right)}
\left(\dfrac{1-q^{2n}}{1+q^{2n}}\right)^{p-2}
\left(\dfrac{1+q^{2n-1}}{1-q^{2n-1}}\right)^{4}
\right]
\nonumber\\
&~&\longrightarrow \left(1+8e^{-2t}+2
\left[(10-p)+{\rm Tr}(M^{T}_{1}M_{2})\right]
e^{-4t}\right).
\label{po}
\end{eqnarray}
Thus, the contribution of the massless states of this
sector to the long-range force is given by
\begin{eqnarray}
\mathcal{A}^{\rm (0)T}_{\rm R-R}=
\frac{T_{p}^{2}\alpha'V_p}
{8(2\pi)^{4-p}}\frac{1}{\mathcal{V}}
\;(\tilde{\xi}+\tilde{\xi'} )
\int_{0}^{\infty}dt~\bigg\{
\Big{(}\sqrt{\frac{\pi}{\alpha' t}}\Big{)}^{4-p}
\exp\left(-\frac{1}{4\alpha't}\sum_{i}
{\left(y_{1}^{i}-y_{2}^{i}\right)^2}\right)\bigg\}.
\label{po}
\end{eqnarray}
As we saw the zero-point energy of the total
Hamiltonian of the twisted R-R sector
is zero. Thus, the ground state of closed superstring
in this sector is massless.
This elucidates that the long-range force (4.6) purely
comes from the exchange of this ground state.
\section{Conclusions and summary}

We constructed two boundary states in
the untwisted and twisted sectors of superstring theory,
corresponding to a fractional D$p$-brane.
The brane lives at the fixed-points of the
orbifold $\mathbb{C}^{2}/\mathbb{Z}_{2}$, and
was dressed by the Kalb-Ramond field and a $U(1)$
internal gauge potential. Besides, transverse
and tangential linear motions and rotation
were imposed to it. We saw that the orbifold
directions, background fields and dynamics
of the brane prominently affected the boundary states.
For example, the orbifold projection induced a
zero-mode part to the twisted NS-NS boundary state.

We observed that the momentum of the emitted closed string
possesses components along the brane worldvolume
and along the direction of the transverse motion.
This noticeable result is unlike the conventional case, and
manifestly originates from the brane dynamics.
Thus, the emitted closed string receives
an effective potential via the brane dynamics.
In fact, having this effect strictly
puts a restriction on the matrix elements
of $\omega_{\alpha\beta}$ and transverse velocity.

We obtained the total interaction amplitude for two parallel
dynamical-fractional D$p$-branes (or a brane and an anti-brane),
in the foregoing setup. The amplitude of the NS-NS part
of the twisted sector does not receive any
contribution from the GSO-projected parts of the
boundary states with different spin structures.
Presence of the various parameters in the setup,
accompanied by the orbifold projection,
gave a generalized form to the total
amplitude. The interaction strength can be accurately
adjusted to any desirable value by these parameters.
However, because of the effects of the
parameters and the orbifold directions
the configuration does not satisfy the BPS
no-force condition.

We separated a special part of the interaction
which merely occurs by the exchange of the massless states of
closed superstring. The untwisted part of this interaction
elaborates exchange of the graviton, dilaton,
Kalb-Ramond and the usual R-R massless states.
The twisted part of the
long-range force specifies exchange of
the NS-NS and R-R ground states which,
in the projected spectrum, are massless. Note that
the spectrum of the projected superstring theory,
by the orbifold, does not comprise the usual
NS-NS and R-R massless states.


\end{document}